# Polarization control of multiply-scattered light through random media by wavefront shaping

Yefeng Guan,[1,2] Ori Katz,[1] Eran Small,[1] Jianying Zhou,[2] Yaron Silberberg,[1]

[1]Department of Physics of Complex Systems, Weizmann Institute of Science, Rehovot 76100, Israel,
[2]State Key Laboratory of Optoelectronic Materials and Technologies, Sun Yat-sen University, Guangzhou 510275, China
(Dated: June 18, 2012)

We show that the polarization state of coherent light propagating through an optically thick multiple-scattering medium, can be controlled by wavefront shaping, i.e. by controlling only the spatial phase of the incoming field with a spatial light modulator. Any polarization state of light at any spatial position behind the scattering medium can be attained with this technique. Thus, transforming the random medium to an arbitrary optical polarization component becomes possible.

The propagation of light in inhomogeneous media, such as biological tissues and the turbulent atmosphere, results in wavefront distortion and scattering. If the light is spatially and temporally coherent, this leads to the formation of speckle patterns [1], which are random spatial intensity fluctuations formed by the distorted wavefront. In addition to the spatial distortions, multiple-scattering also randomly alters the polarization state of the incident light [2], and its temporal and spectral characteristics [3]. However, although multiple-scattering is a random process, it is a deterministic one and its spatial effects can be undone by phase-conjugation [1, 4]. Phase-conjugation is the monochromatic case of the more general technique of time-reversal, which was extensively studied by Fink and colleagues in acoustics and radio-frequency electromagnetic waves [1, 5]. Recently, in optics, it was demonstrated by Vellekoop et al. [6, 7] that effective inversion of the scattering's spatial distortions can be achieved by wavefront-shaping, using a spatial light modulator (SLM) with a surprisingly small number of degrees of control (as compared to the number of scattered modes, i.e. speckles). Following this pioneering work, it was demonstrated that similar techniques can refocus the scattered field in both space and time [8-10], as well as controlling its spectral properties [11]. Interestingly, as predicted two decades ago by Freund [12], these results can be interpreted as the transformation of the random medium to a lens [6-8, 13], a temporal pulse-shaper [8, 9], an arbitrary spectral filter [11] or a mirror [14].

In this letter, we demonstrate the use of a random medium as an arbitrary polarization element. We show that multiply scattered light can be simultaneously refocused and its polarization state controlled by wavefront shaping, i.e. by manipulating only the spatial phase of a linearly-polarized incident beam. The reason that polarization control is achievable by only spatial phase modulation is that multiple-scattering couples the spatial and polarization degrees of freedom. This result is related to the recent utilization of the spatio-temporal/spatio-spectral coupling in multiple scattering media to attain temporal/spectral control through scattering media [5, 9, 11], and for spatial focusing using the temporal degrees of control [10, 15]. Related recent works by Kohlgraf-Owens and Dogariu have shown that a characterized random medium can be used as a spectral-polarimetric analyzer [16]. Here we extend these results to polarization control, rather than characterization or analysis.

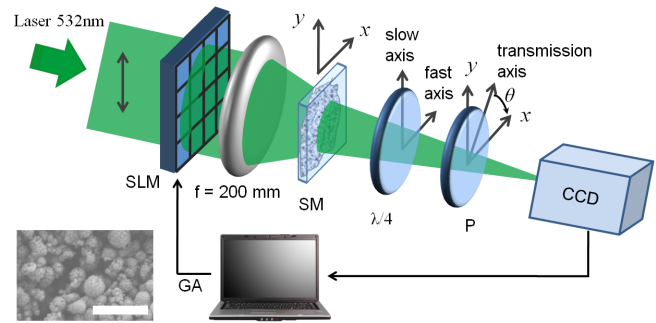

Fig. 1. Experimental setup for controlling the polarization states of scattered light. A wavefront-shaped beam is focused on a multiply scattering medium (SM, a 15-µm-thick $TiO_2$ layer). The scattered light polarization is analyzed and spatially resolved using a polarizer (P) and a CCD camera. A quarter wave plate (QWP) is used to optimize to a circularly polarized state. Inset: SEM image of TiO2 layer. Scale bar, 500nm.

The experimental setup for polarization control by wavefront shaping is shown in Fig. 1. A horizontally-polarized monochromatic collimated beam at 532nm is passed through a phase-only SLM (Hamamatsu LCOS-SLM X10468-02). The phase-shaped light is focused on a strongly scattering sample, which is composed of a 15±5-µm-thick titanium dioxide (TiO2) powder on a 2-mm-thick glass substrate. The average size of TiO2 particles is ~200nm (Fig. 1, inset). The polarization state of the multiply scattered light is analyzed and spatially resolved by a CCD camera (CCD) placed 30cm behind the medium. A rotatable quarter-wave plate (QWP) and a polarizer determine the polarization basis that is imaged on the CCD. Polarization control is achieved by adaptively optimizing the SLM phase-pattern using a genetic algorithm [14] to yield an intensity enhanced spot at the desired spatial position and polarization on the CCD.

When an unshaped (plane-wave) linearly-polarized beam is incident on the random medium, it is scattered to a randomly polarized speckle pattern, losing correlation with its original polarization state (Fig. 2(a-c)). The scattered light has a well defined polarization locally within a coherence area (single speckle), but the polarization of different speckles is uncorrelated [2]. Fig. 2(a) and 2(b) show the typical intensity patterns at the CCD plane for horizontal and vertical polarization components selected by rotating the polarizer, for the unshaped input beam (flat SLM phase pattern). It is evident that a great amount of incident light (Up to 45%) is transferred from the original (horizontal) component to the orthogonal (vertical) component by the scattering medium. The polarization ellipses of three chosen speckles indicated by 'A', 'B' and 'C' in Fig. 2(a) are presented in Fig. 2(c), and demonstrate the random polarized nature of the scattered field (the technique used for the polarization analysis is detailed below). In addition, the two speckle patterns at the orthogonal polarizations are uncorrelated. Similar to previous works, an arbitrary speckle in one of the patterns can be chosen and its intensity enhanced by means of wavefront shaping of the input field [6]. This will result in a brighter spot at a desired location and polarization. However, because the speckle patterns at orthogonal polarizations are uncorrelated, the orthogonal polarization intensity would not be enhanced (Fig. 2(d-h)). Since the choice of polarization basis is arbitrary, this procedure enables obtaining a bright spot at any desired polarization. Thus, wavefront shaping can be utilized to fully control the polarization state at a chosen spatial position behind the medium.

To demonstrate this we have performed several experiments, where in each experiment a target area equal to the average speckle size was selected on the CCD, and its intensity was optimized at an arbitrarily chosen polarization state by setting the QWP and polarizer angles. The intensity optimization was done by a genetic algorithm controlling the incident spatial-phase using the SLM with 80×60 independent segments. Fig. 2(d-l) show the intensity patterns at the two orthogonal polarizations and polarization ellipses for three different experiments, optimizing a single speckle intensity towards a vertical (d-f), horizontal (g-i) and circular (j-l) polarization. We note that the QWP is unnecessary for optimization towards a linear polarization state. For optimizing towards a circular polarization, the fast axis of the QWP was set at 45 degrees to the polarizer. Our results show that, as expected, the intensity of the optimized spot can be enhanced by three orders of magnitude, and the extinction ratio measurement was more than 130:1, which is limited by the polarizer, signal to noise of the CCD, and the optimization time. For the circular polarization optimization, we note that the optimized beam is slightly elliptical due to the imperfect performance of the QWP used.

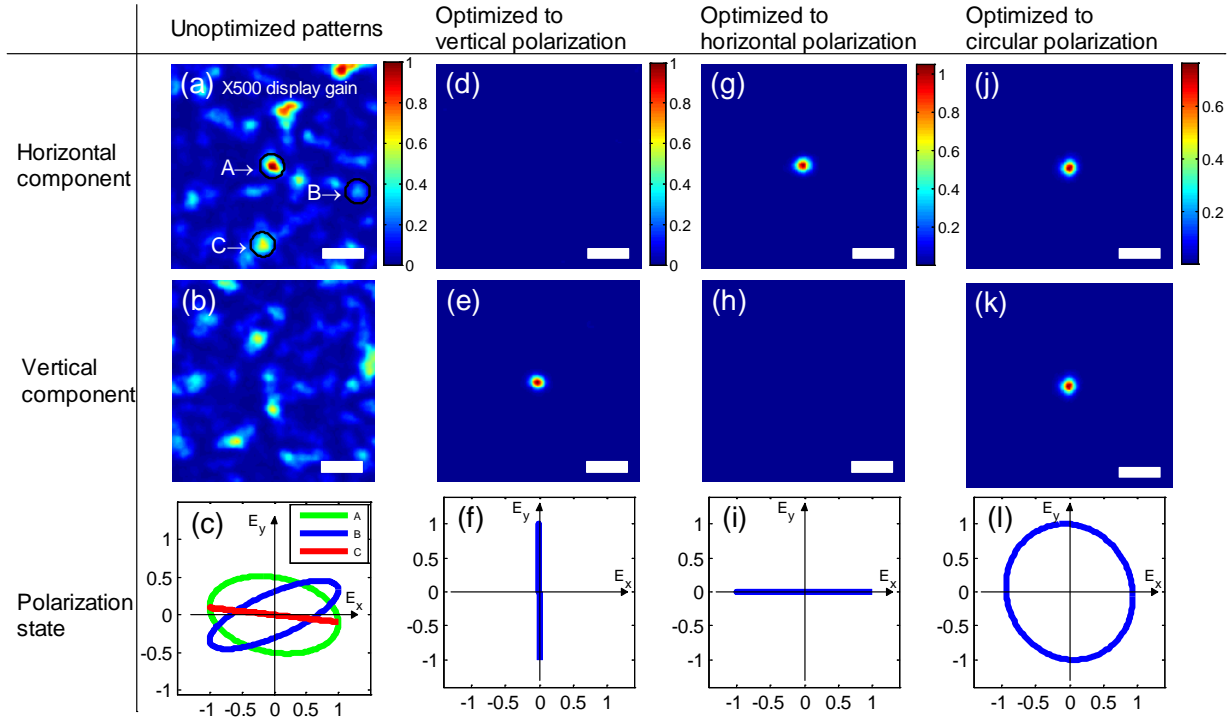

Fig. 2. Measured intensity patterns of the scattered light at horizontal and vertical polarization before and after optimization for three different experiments, optimizing towards vertical, horizontal and circular polarization, and the measured polarization ellipse for a few chosen speckles. At each column (experiment) the images are displayed after normalization with the same factor. (a),(b) Intensity patterns of the scattered light before optimization (flat phase SLM, scale X500 as compared with (c)-(h)). (c) The polarization states of the speckles A, B and C indicated in (a). (d),(e), (g),(h) and (j),(k) The intensity patterns of the scattered light optimized to vertical polarization, horizontal polarization and circular polarization. (f),(i),(l) Polarization states of the optimized beam. Scale bar, 200μm.

The optimized and un-optimized patterns polarization states, as presented in Fig. 2(c, f, i, l) are characterized and spatially resolved in the following way. After removing the QWP, the scattered field is projected at 24 different equally spaced polarizer angles. For each polarizer angle, θ, the CCD image of the intensity pattern is saved $I_\theta(x,y)$. Because the polarization state of every speckle is well defined, its intensity as a function of the polarizer angle behaves according to: $I(\theta)=A[1+\varepsilon\cos(2\theta-\phi)]$, where ε is the polarization ellipticity and φ is the angle of the polarization ellipse's axis (Fig. 3). We found the parameters ε and φ, which are required to draw the polarization ellipses of Fig. 2(c, f, i, l), by fitting $I(\theta)$ with the theoretical cosine function. The raw measured intensity curves for three speckles in the unoptimized case are presented in Fig. 3(a), and the raw measured intensities for the optimized cases are presented in Fig. 3(b). In Fig. 3(b) we have also plotted the best performance of the QWP used to transform a linearly polarized light to a circularly polarized one without the scattering medium, verifying that our optimization results are limited by the performance of the waveplate used. This is by no means a limit to the presented technique.

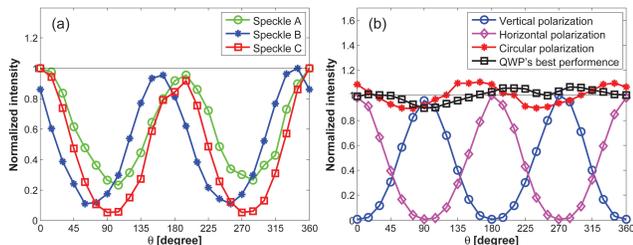

Fig. 3. (a) The intensity measurement of the speckles A, B and C in Fig. 1(a) by rotating the polarizer. It shows that these speckles are in the different polarization states. (b) The intensity measurement of the optimized light beam for the vertical, horizontal and circular polarization in Fig. 1(d-l), and the QWP's best performance is also shown. All intensities are normalized to arbitrary units, indicated by the black solid line.

Finally we show that similar to previous results in wavefront shaping, the technique is not limited to controlling a single spatial position, and several speckles at different spatial positions can be optimized simultaneously to an arbitrary chosen polarization state (Fig. 4). In this case the enhancement for each optimized beam will be reduced by a factor equal to the number of speckles that are optimized.

In conclusion, we have demonstrated the control of polarization states of multiply scattered light by wavefront shaping. We have used an adaptive closed-loop optimization algorithm to find the desired phase pattern. Alternative methods such as transmission matrix [13], can be used. Using the demonstrated technique, one can dynamically generate highly complicated intensity and polarization patterns which can prove useful for polarization sensitive sensing and imaging, either in complex media, or by exploiting a complex medium as the optical instrument for generating the complex interrogating beams [12]. Combining the presented technique with the recently demonstrated schemes for temporal [8, 9] and spectral control [11], it will allow full control of multiply scattered light through a scattering medium in space, time, frequency and polarization, by wavefront shaping alone.

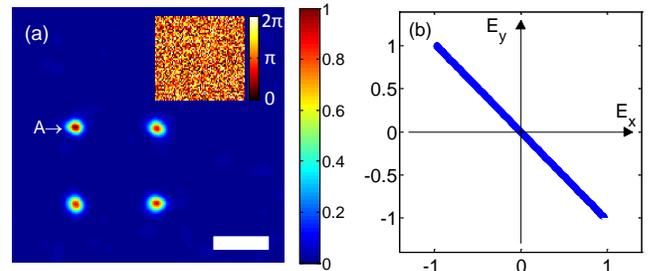

Fig. 4. (a) Simultaneous optimization for four linearly polarized speckles. The four speckles were all optimized to the same polarization state at 45 degrees to the input polarization. Inset, the phase pattern put on the SLM to form (a). (b) The measured polarization ellipse of the speckle marked as 'A' in (a). Scale bar, 200μm.

We thank S. Izhakov for providing and analyzing the scattering sample. Y.G. is supported by the Oversea Study Program of Guangzhou Elite Project. E.S. is supported by the Adams Fellowship Program of the Israel Academy of Sciences and Humanities. This work was supported by MINERVA and the ERC grant QUAMI.